\documentclass[newstyle,twocolumn,proceedings]{rmaa}

\title{VIIZw 403: a BCD with a noncoeval starburst}

\suppressfulladdresses
\author{S. Silich \altaffilmark{1,2}, G. Tenorio-Tagle  \altaffilmark{1},
        C. Mu\~noz-Tu\~non \altaffilmark{3} and 
        L. M. Cairos \altaffilmark{3}}
\altaffiltext{1}{Instituto Nacional de Astrof\'\i sica Optica y Electr\'onica, 
AP 51, 72000 Puebla, M\'exico}
\altaffiltext{2}{Main Astronomical Observatory National Academy of Sciences of
Ukraine, 03680, Kiev-127, Golosiiv, Ukraine}
\altaffiltext{3}{Instituto de Astrofisica de Canarias, E 38200 La Laguna, 
Tenerife, Spain}

\shortauthor{Silich}
\shorttitle{VIIZw 403}

\listofauthors{S.~Silich}
\indexauthor{Silich, S.}

\begin{document}
\maketitle

\noindent\textbf{The recent history of star formation in VI~IZw 403
is analyzed taking into  account the dynamics of the starburst blown 
superbubble  
and the restrictions that follow from X-ray observations, and from 
our H$\alpha$ data. Our results show that the starburst energizing 
VIIZw 403 is NOT coeval, but rather the star forming phase has lasted for 
more than 30Myr
and at such a low pace that most probably the newly processed metals will not 
be ejected into the intergalactic medium.}

VII Zw403 is a BCD galaxy, in recent years considered in many 
discussions related to star formation histories and the possible
impact of dwarf systems on the surrounding intergalactic medium.
It is an isolated system, however considered to be a member
of M81 group.  The X-ray observations
(Papaderos et al. 1994; Fourniol 1997) added more interest to the
system, as they revealed an extended kpc-scale region of diffuse
X-ray emission. 

Several models have been investigated with the aim of matching the 
main observed parameters of VIIZw 403: The H$\alpha$ luminosity 
($L_{H\alpha} = 1.8 \times 10^{39}$ erg s$^{-1}$),
the X-ray power ($L_{X} = 2.3 \times 10^{38}$ erg s$^{-1}$) and the 
size of the diffuse X-ray emitting region ($\sim 1$kpc). 
In all of them we assumed 
that the photons presently produced by the stellar clusters are all
used up to reestablish the ionization of the central HII region.
This fact is supported by the HI mass present in VIIZw 403.
The ${H\alpha}$ flux defines then the present star formation rate 
which we have assumed to be constant in time. 

We have further assumed that the observed HI mass occupies a smooth low 
density halo but an important fraction  of it is in a dense cloud component 
($M_{cl}$). This has a major impact on the evolution as it affects both the 
time required to reach a given size as well as the X-ray luminosity 
produced by the superbubble.

The calculations were carried out with our 3D Lagrangian code,
which accounts for the enrichment of the hot superbubble interior
by the metals ejected via supernova explosions (Silich et al. 2001).
\begin{figure}[!t]
\includegraphics[width=\columnwidth]{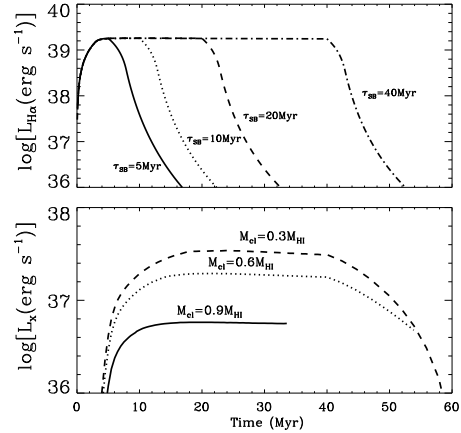}
  \caption{The calculated central star cluster $H_{\alpha}$ luminosity 
(a) for starburst of different duration, and bubble X-ray emission for 
different mass fractions in the assumed cloud component (b) 
as functions of time}
    \label{fig1}
\end{figure}
The results from the calculations imply (see figure 1) that:

- The starburst energizing VIIZw 403 is NOT an instantaneous or
  coeval star cluster.

-The starburst formation time is larger than 30Myr. 
 This is also the time span required for the 
 starburst blown superbubble to reach the dimension and
 luminosity of the extended X-ray component, while producing the
 observed  ${H\alpha}$ luminosity. 

- During that time star formation has proceeded at an almost
  constant rate SFR = $4 \times 10^{-3}$ M$_{\odot}$ yr$^{-1}$.

- The combination of the various parameters used to produce 
  the  {\it best} model imply that the newly processed metals
  in VIIZw 403 will not be ejected into the intergalactic medium.

\end{document}